\documentclass[a4paper]{article} 

\usepackage{url,hyperref,lineno,microtype,subcaption}
\usepackage[onehalfspacing]{setspace}
\usepackage{multirow}
\usepackage{natbib}
\usepackage{graphicx}
\usepackage{fancyhdr}

\begin{document}

\title{\textbf{Tropical Cyclone Track Forecasting using Fused Deep Learning from Aligned Reanalysis Data}} 

\author{Sophie Giffard-Roisin\,$^{1,2,*}$, Mo Yang\,$^{3}$, Guillaume Charpiat\,$^{4}$,\\
 Christina Kumler Bonfanti\,$^{5}$, Balázs Kégl\,$^{3}$, and Claire Monteleoni\,$^{1}$}
\date{\small{$^{1}$Computer Science Department, University of Colorado, Boulder, Colorado, USA \\
$^{2}$Univ. Grenoble Alpes, Univ. Savoie Mont Blanc, CNRS, IRD, IFSTTAR, ISTerre, Grenoble, France.  \\
$^{3}$Linear Accelerator Laboratory, Université Paris-Sud, CNRS, France\\
$^{4}$Inria Saclay–Ile-de-France, LRI, Université Paris-Sud, France\\
$^{5}$ University of Colorado Boulder, Cooperative Institute for Research in Environmental Sciences, NOAA/OAR/ESRL/Global Systems Division, Boulder, Colorado, USA.\\}
\textit{Corr. author: sophie.giffard@univ-grenoble-alpes.fr} }

\maketitle
\thispagestyle{fancy}

\begin{abstract}

The forecast of tropical cyclone trajectories is crucial for the protection of people and property. Although forecast dynamical models can provide high-precision short-term forecasts, they are computationally demanding, and current statistical forecasting models have much room for improvement given that the database of past hurricanes is constantly growing. Machine learning methods, that can capture non-linearities and complex relations, have only been scarcely tested for this application. We propose a neural network model fusing past trajectory data and reanalysis atmospheric images (wind and pressure 3D fields). We use a moving frame of reference that follows the storm center for the 24h tracking forecast. The network is trained to estimate the longitude and latitude displacement of tropical cyclones and depressions from a large database from both hemispheres (more than 3000 storms since 1979, sampled at a 6 hour frequency). The advantage of the fused network is demonstrated and a comparison with current forecast models shows that deep learning methods could provide a valuable and complementary prediction. Moreover, our method can give a forecast for a new storm in a few seconds, which is an important asset for real-time forecasts compared to traditional forecasts.

\end{abstract}

\section{Introduction}

Cyclones, hurricanes and typhoons are words designating the same phenomena: a rare and complex event characterized by strong winds surrounding a low pressure area. The ability to forecast their trajectory and intensity forecasts is crucial for the protection of people and property. However, their evolution depends on many factors at different scales, altitudes and times, which leads to modeling difficulties \citep{emanuel2003tropical}. As the dynamical models evolve, their forecast accuracy improves; however, historical tropical cyclone databases have scarcely been utilized by machine learning and deep learning methods, to further improve forecast accuracy.
\\

\subsection{Existing Storm Forecasts Methods}

Today, the forecasts (track and intensity) are provided by numerous guidance models\footnote{NHC track and intensity models, \url{https://www.nhc.noaa.gov/modelsummary.shtml}, Accessed: 2018-07-04.}. Dynamical models solve the physical equations governing motions in the atmosphere and they are influenced by physical models -convective schemes (such as Kain-Fritsch or Simplified Arakawa Schubert), cloud microphysics, land surface model, ocean model, sea/land ice model, planetary boundary layer scheme, surface layer scheme, longwave and shortwave radiation schemes, subgrid-scale diffusion- and by their data assimilation methods (such as 4D-VAR). They are computationally demanding and in current practice older model runs are adjusted in order to be considered \emph{early} methods, i.e. available in real time. Current forecasts produced by regional specialized meteorological centers, like the American Official NHC Forecast (OFCL), are driven by consensus or ensemble methods able to combine different dynamical models \footnotemark[1] (up to 20 models for the Global Ensemble Forecast System \footnote{GEFS, \url{https://www.ncdc.noaa.gov/data-access/model-data/model-datasets/global-ensemble-forecast-system-gefs}, Accessed: 2018-12-12.}). Statistical models, in contrast, are based on historical relationships between storm behavior and various other parameters (\citet{demaria2005further}). However, they are based on simple regressions on few statistical features. By incorporating large spatial atmospheric data in a statistical model, using state-of-the-art machine learning methods, we can improve the accuracy while reducing the calculation time.
\\

\subsection{Deep Learning and Convolutional Neural Networks (CNN)}

A convolutional neural network (CNN) is a deep learning architecture widely adopted as a very effective model for analyzing images or image-like data for pattern recognition \citep{krizhevsky2012imagenet, milletari2016v}. A CNN is structured in layers: an input layer connected to the data, an output layer connected to the quantities to estimate, and multiple hidden layers in between. The hidden layers of a CNN typically consist of convolutional layers, pooling layers, fully connected layers and normalization layers. The convolutional operations are inspired by the cortex visual system, where each neuron only processes data for its receptive field. Fully connected (FC) layers, usually at the end of the network, connect every neuron in one layer to every neuron in another layer. The advantage of CNN is that it can learn to recognize spatial patterns by exploiting translation invariance (i.e. all parts of the image are processed in a similar way), and thus can extract features automatically while considerably reducing the number of parameters. 
 
Recurrent neural networks (RNN) are a class of artificial neural networks that can model temporal dynamic behavior for a time sequence. Unlike feedforward neural networks (like CNNs), RNNs can use their internal state (memory) to process sequences of inputs. An LSTM (long short term memory) network is a particular RNN used in different time-series applications. Even though the long-short-term memory (LSTM) networks are among the most successful methods for predicting time-series events, they are still difficult to train, and simpler CNNs may outperform LSTMs (\citet{bai2018empirical}). Moreover, encoding time frames as different input channels in a CNN architecture already proved its efficiency if the history size is fixed (\citet{de2017deep}). In our forecast problem we are dealing with spatial and time-varying meteorological data, with a short history size needed (few time steps): the CNN encoding time frames as different input channels is more suited than the LSTM model.
\\

\subsection{Machine Learning and Deep Learning in Forecasting Problems}

Current statistical forecasting models still perform poorly with respect to dynamical models, even though the database made of past tropical cyclones is constantly growing \footnote{NHC track and intensity models, https://www.nhc.noaa.gov/verification/verify6.shtml, Accessed: 2018-07-04.}. Machine learning methods, which are able to capture non-linearities and complex relations, have only been scarcely tested for tropical cyclone tracking. Yet, they have recently shown their efficiency in a number of various other forecasting tasks. In particular, CNNs have raised attention as they are suited for large imaging (2D or 3D) data. In \citep{xingjian2015convolutional}, a convolutional LSTM model was used for precipitation forecast. Another recent study predicts the evolution of sea surface temperature maps by combining CNNs with physical knowledge (\citet{de2017deep}). CNNs have also been used for the detection of extreme weather events like tropical cyclones from weather model variables such as integrated water vapor, as in \citet{racah2017extremeweather} or in \citet{kim2019deep}. These studies show that inputs from meteorological fields are suited for training CNN models in various forecastings and detection problems.

However, only few preliminary studies have tackled tropical cyclone forecast tracking using machine learning. Two studies used RNN from only trajectory information. \citet{moradi2016sparse} was tested on 6h- and 12h-forecast on only 4 tropical cyclones, while \citet{gao2018nowcasting} was tested on only Northwest Pacific tracks. Yet, the trajectory data information is limited so the use of local meteorological fields is crucial for this complex problem. \citet{ruttgers2018prediction} proposes to use a generative adversarial network (GAN) to generate the future RGB atmospheric image (harder problem), but only for a 6 hour prediction. The GAN method is interesting, but the forecast errors are large. To be noticed, the authors are suggesting using in the future some meteorological fields such as the velocity to improve the results. Other studies uses storm tracks and reanalysis maps as input for a hybrid CNN - LSTM network in order to learn the (x,y) tracking coordinates (\citet{mudigonda2017segmenting, kim2019deep}). While these methods are usually not compared with existing forecasts methods, some of them seem to even perform worse than a baseline of constant speed and direction, see \citet{giffardroisin:hal-01851001}. Based on this review and to the best of our knowledge, recent efforts have been made in developing innovative machine learning forecast methods but the results are still poor, the forecast time is always less than 15 hours and a good evaluation is usually lacking.
\\

\subsection{Frame of Reference}
\begin{figure}[t]
\centering
\includegraphics[width=0.85\hsize]{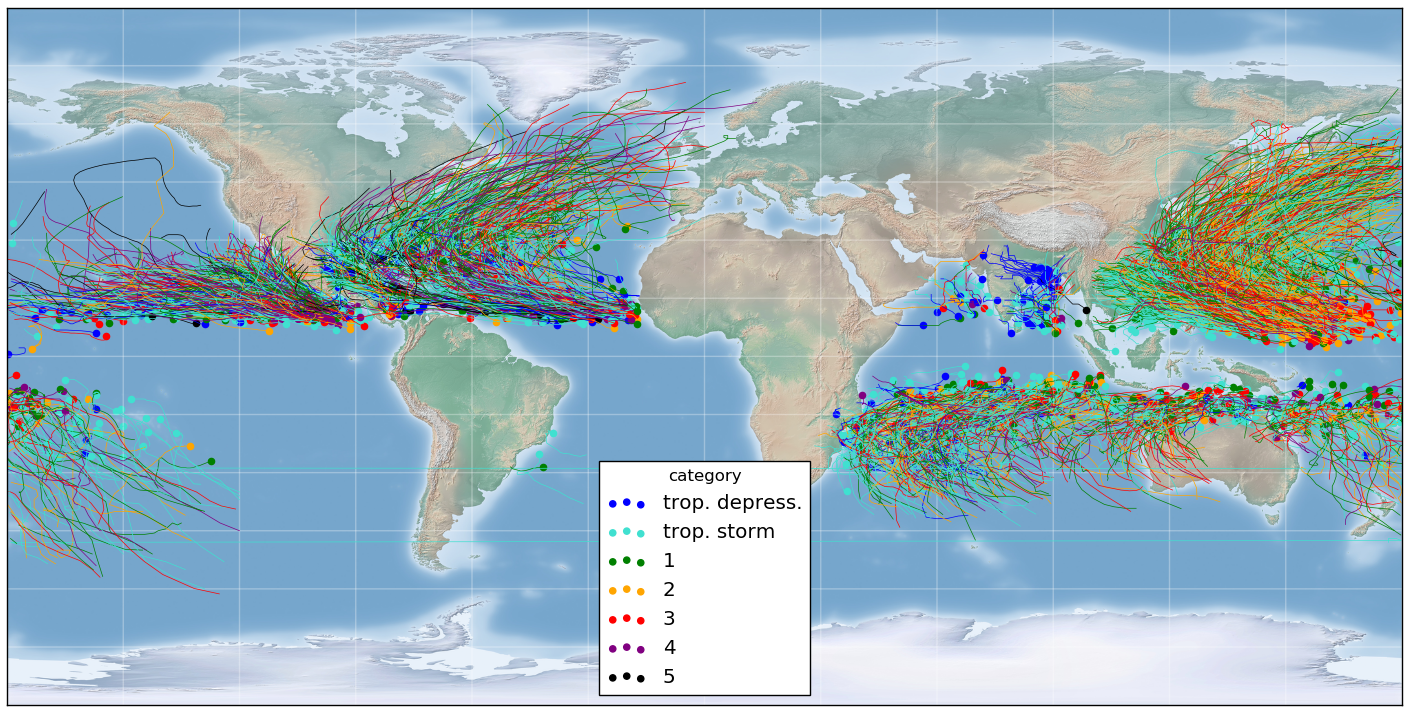}
\caption{Database: more than 3000 tropical/extra-tropical storm tracks since 1979. Dots = initial position, colors = maximal storm strength according to the Saffir-Simpson scale.}
\label{fig:earth}
\end{figure}

In order to overcome these limitations, we developed a different strategy to use image-like data as inputs for training a deep learning network.
All of the current machine learning forecasts consider a fixed regional map for tracking storms, of size 160 x 80 deg (longitude/latitude) for \citet{mudigonda2017segmenting} and \citet{kim2019deep} and of the size of the Korean peninsula area (around 30 x 30 deg) for \citet{ruttgers2018prediction}. However, a fixed region for tropical cyclone forecast has three major limitations. First, the tracked storm must stay in the region even though tracks often cross oceans (see Figure \ref{fig:earth}), forcing the uses of a large region, even if it leads to memory issues (\citet{mudigonda2017segmenting}). Moreover, learning local phenomena on a large and non-centered image can be difficult. Finally, it prevents information transfer between storms coming from different basins or regions, where ground truth data is scarce. In our recent work (\citet{giffardroisin:hal-01851001}), we showed the advantage of using a moving reference CNN model for forecasting tropical cyclone tracks 6 hours into the future. This gave roughly a 30km mean error whereas all other learning methods have a mean error larger than 60km \citep{moradi2016sparse, ruttgers2018prediction, kim2019deep} and a constant speed baseline gave 46km mean error. However, a 6h-forecast is of little use for catastrophe planning and it is not possible to compare to existing forecast methods as the smallest standard is 24 hours. 
\\
\\

\subsection{Contributions}

In this work, we propose to answer the following questions: Can we develop a statistical forecast model, using state-of-the-art deep learning techniques, that is able to compete with current forecast models at a 24h time lag? How can we take the best advantage of the worldwide historical track database and the reanalysis meteorological fields?

We propose to extend our previous work by using a moving frame of reference that follows the storm center for a 24h-forecast tracking task. We pose the tracking problem as the estimation of the displacement vector, $\vec{d}$, between current and future locations. Moreover, we propose to use the reanalysis data as cropped images (25 x 25 degrees) centered on the storm location. That way, the computation time is reduced and we can infer information from storms coming from a large number of tropical cyclone basins from both hemispheres. In particular, our database is made up of slightly more than 3000 storms since 1979, sampled at a 6 hour frequency (more than 90 000 time steps). We include past temporal information by adding the reanalysis maps from previous time steps. We propose a fusion convolutional neural network taking into account past trajectories and different fields from reanalysis images (wind fields and pressure). Using every time step of a storm as a unique sample (thus having 90 K samples) allows us to train CNN algorithms that require big data to optimize their large number of parameters (here of the order of $10^{6}$). This paper focuses on a 24h-forecast as a proof of concept, and could be easily extended to larger forecast times.

\begin{figure}[t]
  \centering
  \includegraphics[width=0.95\hsize]{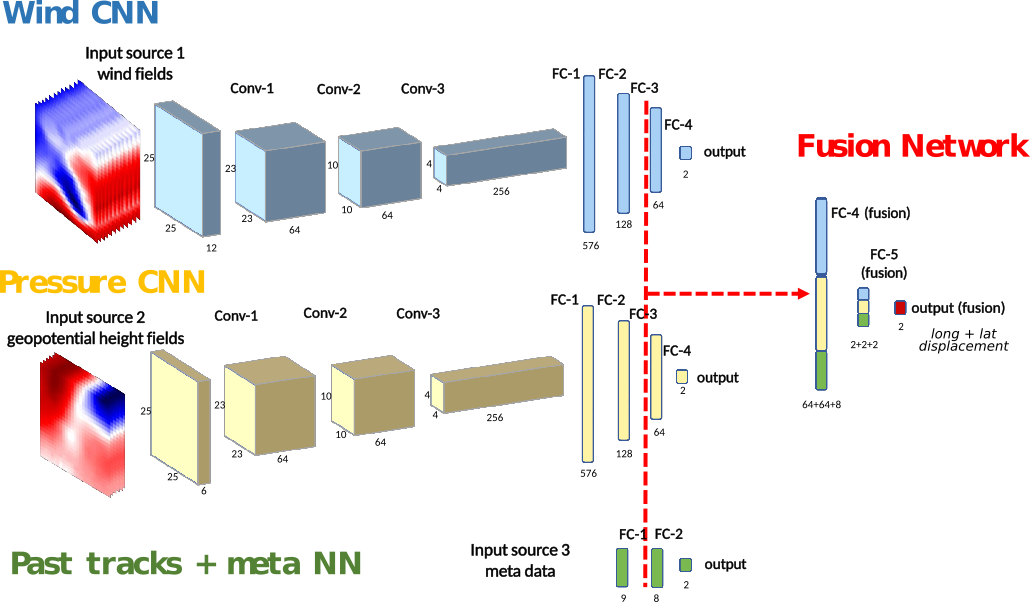}
  \caption{General architecture: the three types of data feed three neural networks trained separately. The final fused network is re-trained before predicting the 24h-forecast displacement.}
    \label{fig:fusion}
\end{figure}

We aim at building an end-to-end model using two types of data (track data and 3D reanalysis) as input. For each time step of each storm, we want to independently estimate its future displacement. After presenting the data, we will show how we designed CNNs to learn from the reanalysis and then improved the result by combining it with history tracks and other 0D features (such as longitude, latitude, and maximal sustained windspeed). Figure \ref{fig:fusion} summarizes the fusion pipeline that predicts the 24h storm displacement. Lastly, we will show the results on the test set and compare these with current forecast models.
\\

\section{Data Description}

\label{sec:Data}

\subsection{Storm Tracks}

The raw storm track data used in this study is composed of more than 3000 tropical and extra-tropical storm tracks since 1979, extracted from the NOAA database IBTrACS (International Best Track Archive for Climate Stewardship, \citet{knapp2010international}), shown in Figure \ref{fig:earth}. The tracks were produced by multiple governmental agencies, depending on the basin. They are defined by the 6-hourly center locations (latitude and longitude), and the database also includes some associated descriptors such as the windspeed (see Section \ref{sec:features}). It includes both hemispheres and the number of records per storm varies from 2 to 120 time steps. In total, the database counts more than 90,000 time steps and we used our method to predict the 24-h track forecast for each single time step.
\\

\subsection{Reanalysis Data}

The trajectory of a storm depends on large-scale atmospheric flows. We chose to extract analyzed atmospheric fields from reanalysis data, not the forecast fields. 
We used the ERA-Interim database (\citet{dee2011era}), which is one of the reanalysis datasets covering the data-rich period since 1979.
Reanalysis is a systematic approach to produce datasets for climate monitoring and research,
covering the entire globe from the Earth’s surface to well above the stratosphere and estimate hundreds of available variables.
 ERA-Interim is a global atmospheric reanalysis produced by the European Centre for Medium-Range Weather Forecasts (ECMWF) and is produced in near to real time. The spectral resolution is T255 (around 80 km), the time resolution is 6 hours, and there are 60 vertical pressure levels until 0.1 hPa (altitude around 64 km).
\\

\subsection{Feature Selection}

\label{sec:features}
In this work, we used storm track data and reanalysis outputs to forecast tropical cyclone tracks. We can classify them into 4 types of information:

\begin{itemize}

\item Past displacements (1D). We define a displacement as the values $(\delta long_{\Delta t},\delta lat_{\Delta t})$ between the locations of a storm's center, as recorded in the storm track data, at different times. The time difference, $\Delta t$, being in a multiple of 6 hours. The historical displacements of a storm help predict its future displacement $(\delta long_{24h},\delta lat_{24h})$. We used the current displacement (i.e. between times $t - 6h$ and $t$) and the past displacement (between $t - 12h$ and $t - 6h$).  These features are 1D in the sense that they are defined for each past time step (1D temporal data).

\item \textbf{Meta data (0D)}. We use all of the features extracted from the IBTrACS database, as they contain crucial information related to the TCs: the current center-point latitude and longitude, the current windspeed at the center of the storm, the current distance to land, and the Jday predictor (Gaussian function of \textit{Julian day of storm init - peak day of the tropical cyclone season in the hemisphere}, see \citet{demaria2005further}). We refer to such features as 0D because they are not defined on a spatial grid.

\item \textbf{Wind fields $u$ and $v$ (spatial fields, 3D).} We applied a sparse feature selection technique (Automatic Relevance Determination, based on linear regression to the target displacement shift) over the 10 available reanalysis fields on pressure levels, which highlighted the usefulness of two reanalysis fields in particular: wind fields and the geopotential height. Wind fields are the direct observations of the atmospheric flows, so their importance is clear. In order to have a moving frame of reference, we extracted the wind fields of the neighborhood of the storm at every time step from the ERA-interim reanalysis database, see Figure \ref{fig:hurr_schema}. Specifically, we extracted the u-wind and v-wind fields on a 25x25 degree grid centered on the current storm location, at three atmospheric pressure levels (700 hPa, 500 hPa, and 225 hPa). The choice of the three pressure levels was inspired by the literature on statistical forecast models (\citet{demaria2005further}) a well as on a sensitivity analysis. The idea is to capture the air movements at the different levels of the troposphere: low, mid and high clouds. The size of the grid was motivated by the fact that meaningful information is extractable from the movement of air masses around the storm, while focusing on the storm only (10x10 deg) bounds most historic extreme storms (which was confirmed by preliminary training experiments).

\begin{figure}[t]
  \centering
   \includegraphics[width=0.5\hsize]{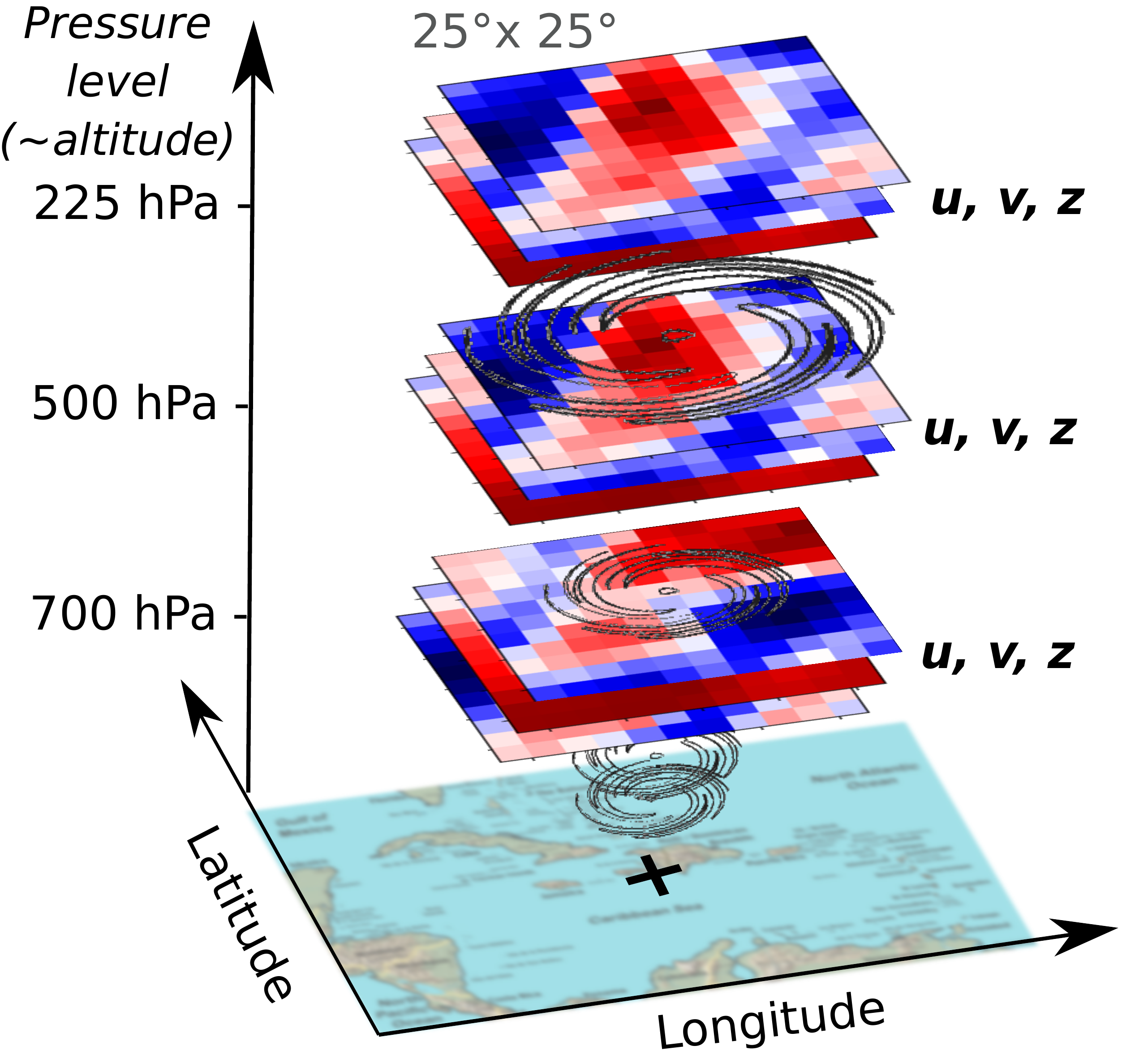}
  \caption{Global atmospheric grids centered on the storm location: wind fields (u and v) and geopotential height (z).}
  \label{fig:hurr_schema}
\end{figure}

\item \textbf{Geopotential height fields $z$ (spatial fields, 3D).} As previously mentioned, the geopotential height was also found relevant for this task from the ARD regression. 
Similar to wind fields, we extracted the geopotential height (or iso-pressure latitude) fields of the neighborhood of the storm at every time step on a 25x25 degree grid centered on the current storm location, at three atmospheric pressure levels (700 hPa, 500 hPa, and 225 hPa).

\end{itemize}

In order to capture the dynamics, we extracted the wind fields and the geopotential height measured at times $t$ and $t-6h$ at the same location. These fields are thus 3D (spatial) x 1D (temporal). We point out that we first used surface reanalysis data, including sea surface temperature, sea level pressure and 10 meter winds, but because of no significant impacts to the result, we concentrated our efforts to atmospheric wind and geopotential fields.
\\

\subsection{Set Separation}

The storms were randomly separated in three sets as following: training (60\%) / validation (20\%) / testing (20\%). Thus, the storms in the test set have never been seen before by the learning algorithm.
Then, within each set, all time instants were treated independently. The training set was used for optimizing the parameters of the neural networks (back-propagation). The validation set was used to select the architecture of the network (Section \ref{Methodology}). Finally, the test set was kept hidden and was only used to show the final prediction accuracy at test-time (Section \ref{results}).

\section{Methodology: a Deep Fusion Model}

\label{Methodology}

\subsection{Overview}

Because of the differing nature of the data sources, it is not straightforward to mix all the data into a neural network (NN); different learning rates are needed. We propose a new fusion NN architecture taking into account the four sources of information. An overview of the architecture we developed is shown in Figure \ref{fig:fusion}. We divided our fusion architecture into three branches: a Wind CNN, a Pressure CNN and a Past tracks + meta NN. The Wind CNN and Pressure CNN are 2D CNNs that take atmospheric fields (long, lat, stacked over height and time) as input, while the Past tracks + meta NN is a small neural network which takes 0D features as input (stacked over time). Each branch of the network makes its predictions independently. We train the parameters of each individual branch of the network for the same task, i.e. predicting the 24h-forecast track. We then integrate the three networks into a fused network and fine-tune the parameters. The different steps will be outlined in the following sections.
\\
\\

\subsection{Convolutional Neural Network for Reanalysis}

The reanalysis data are 2D fields on a grid of fixed size. Treating them as images has the advantage to give access to a large literature on image processing, where CNN is the current state-of-the-art.
We propose two similar CNN networks for the wind and the pressure fields. We separate them into two networks because the type of data is different and thus different learning rates were needed. We stacked the data over height (pressure level) and time, such that the inputs of the CNNs consist of multiple 2D (long, lat) frames or channels. 
The Pressure CNN has six input channels (each one of size 25x25), while the Wind CNN input consists of 12 channels ($u$ and $v$ are stacked). We used a typical CNN architecture, alternating convolutional layers (Conv layer) and max-pooling layers, with fully connected layers at the end \citep{simonyan2014very}. Following conventional wisdom in the computer vision literature, all hidden layers are equipped with the rectification (ReLU) non-linearity and batch normalization. In order to select the best architecture, the different configurations that we have evaluated for Wind CNN and Pressure CNN are outlined in Table \ref{tab:vgg_arch}, one per column. All configurations follow the generic design described above and differ only in depth, which is determined by the number of convolutional layers. As shown in Table \ref{tab:neuron_nums}, in order to have fair comparisons among the architectures, we designed configurations with approximately the same number of parameters to estimate.

\begin{table}[t]
	\begin{center}
		\caption{Different configurations of \textit{Wind CNN} tested. The depth of the configuration increases from left to right, as more layers are added. \emph{conv3-32} indicates a convolution of size 3x3 with 32 output features. \emph{FC} means fully connected layer. \textit{maxpool} indicates a 3x3 max-pooling layer. The ReLU activation and batch normalization layers (applied after each conv. or FC layer) are not shown in the figure.}	\label{tab:vgg_arch}
		\begin{tabular}{|c|c|c|c|}
			\hline
			\multicolumn{4}{|c|}{\textbf{ConvNet Configurations}}                                                                                                                                                                                                                                                                                                                       \\ \hline
			\textbf{A}   & \textbf{B}  & \textbf{C}   & \textbf{D                                                                                                 } \\ \hline
			\textbf{7 layers} & \textbf{8 layers}  & \textbf{9 layers}   & \textbf{10 layers}                                                                                                        \\ \hline
			\multicolumn{4}{|c|}{\textit{input (12 channels of size 25*25)}}                                                                                                                                                                                                                                                                                                               \\ \hline
			\begin{tabular}[t]{@{}c@{}}conv3-32\\ \textit{maxpool}\end{tabular} & \begin{tabular}[t]{@{}c@{}}conv3-32\\ conv3-32\\ \textit{maxpool}\end{tabular} & \begin{tabular}[t]{@{}c@{}}conv3-64\\ conv3-64\\ \textit{maxpool}\\ conv3-256\end{tabular} & \begin{tabular}[t]{@{}c@{}}conv3-64\\ conv3-64\\ \textit{maxpool}\\ conv3-128\\ conv3-256\\ \textit{maxpool}\end{tabular} \\ \hline
			\multicolumn{4}{|c|}{FC-576}                                                                                                                                                                                                                                                                                                                                         \\ \hline
			\multicolumn{4}{|c|}{FC-128}                                                                                                                                                                                                                                                                                                                                         \\ \hline
			\multicolumn{4}{|c|}{FC-64}                                                                                                                                                                                                                                                                                                                                          
\\ \hline
			\multicolumn{4}{|c|}{FC-2}                                                                                                                                                                                                                                                                                                                                          \\ \hline
		\end{tabular}
		
	\caption{Number of parameters (in millions) of the four network configurations tested in Table \ref{tab:vgg_arch}.} 	\label{tab:neuron_nums}
	\begin{tabular}{|l|c|c|c|c|}
		\hline
		Network              & \textbf{A}    & \textbf{B }   & \textbf{C}    & \textbf{D}    \\ \hline
		Number of parameters (x $10^{6}$) & 2.27 & 2.33 & 2.75 & 2.67 \\ \hline
	\end{tabular}		
		
	\end{center}

\end{table}

We evaluated the performance on 24-hour storm track prediction for the Wind CNN. The result of the architecture evaluation on the validation set is shown in Table \ref{tab:arch_evaluation}. We give two scores: Root Mean Square Error (RMSE) and Mean Absolute Error (MAE), in kilometers. With the increase of model depth, there is  no clear improvement on the result. Since adding more convolutional layers allows the network to learn features at more levels of abstraction, we chose the intermediate Network C, which consists of 3 convolutional layers and one maxpooling, followed by 4 fully connected layers.

\begin{table}[]
	\centering
	\caption{Performance of candidate configurations (Wind CNN) on 24 hours storm track prediction, on the validation set using wind fields.}
	\label{tab:arch_evaluation}
	\begin{tabular}{|c|c|c|}
		\hline
		\textbf{Model} (from Table. \ref{tab:vgg_arch}) & \textbf{Root Mean Square Error} ($km$) & \textbf{Mean Absolute Error}($km$) \\ \hline
		\textbf{A} & 177.2 & 145.4  \\ \hline
		\textbf{B} & 178.2 & 146.6  \\ \hline
		\textbf{C} & 177.6 & 145.6 \\ \hline
		\textbf{D} & 178.2 & 146.7 \\ \hline
	\end{tabular}
\end{table}

We also evaluated how adding more historical features from past time steps in the input data can improve performance. In addition to $t$ and $t - 6h$, we did not observe any noticeable improvement by including more data from the same location at previous time steps. We thus only kept the times $t$ and $t - 6h$.
\\

%
%

\subsection{Past tracks + meta Neural Network}


Another important source of information are the previous displacements and the other IBTrACS features (see Section \ref{sec:Data}). We chose to use all of the available features, which can be treated as a size-9 vector of 0D components. We designed a small NN of two small fully connected layers (the green branch in Figure \ref{fig:fusion}) to learn the future track from the 0D features. For such 0D features, we also tested other machine learning methods (support vector regression and random forests), giving similar results. We kept the neural network for a better compatibility with the CNNs. We use two past displacements from $t-12h$ to $t-6h$ and from $t-6h$ to $t$ because more past tracks did not improve the performance. 
\\

\subsection{Fused Neural Network for Wind, Altitude, and Tracks}


Because of the differing natures between the wind fields, pressure fields and past track data, it is not straightforward to mix them as an input of a NN. Indeed, our preliminary experimentation on training a network combining these three types of inputs simultaneously did not give satisfactory enough results. Instead, we first train separately the three individual branches of the network. We then concatenate their two last layers and add a layer at the end of the network (see Figure \ref{fig:fusion}).
The new concatenated layers consist of the same weights as before in each branch, plus new connections from each branch to the other ones, which we initialize to zero. That way, the function computed is (at start) the same as previously. 
We then re-train the whole fused network by allowing every weight to be re-optimized. The number of fused layers (here two) was determined by comparing four different configurations on the validation set, and a different learning rate was tuned on the validation data set for this final optimization.
\\

\subsection{Algorithmic Details}

We trained our networks using the root mean square error (RMSE) in kilometers between the forecast and the true storm location at $t+24h$ as the loss function. We used the Haversine formula, giving great-circle distances between two points on a sphere from their longitudes and latitudes. We added a regularization term in the loss as an L2 penalty on the weights of the model in order to avoid overfitting. The tuning parameter was optimized empirically on the validation set to $0.01$. The training was performed by the Adam optimizer, and each model converged within 200 epochs. Every evaluation was repeated three times and an average score was computed in order to assess the robustness to the random weights initialization. Although the training takes nearly 8 hours using PyTorch 4.0 on 4 TitanX GPUs with data parallelism (\citet{krizhevsky2014one}), the testing time or inference is only a few seconds.
\\

\section{Experimental Evaluation}

\label{results}

\subsection{Results on the Whole Dataset (all basins)}

\begin{figure}[t]
  \centering
		\includegraphics[width=0.7\hsize]{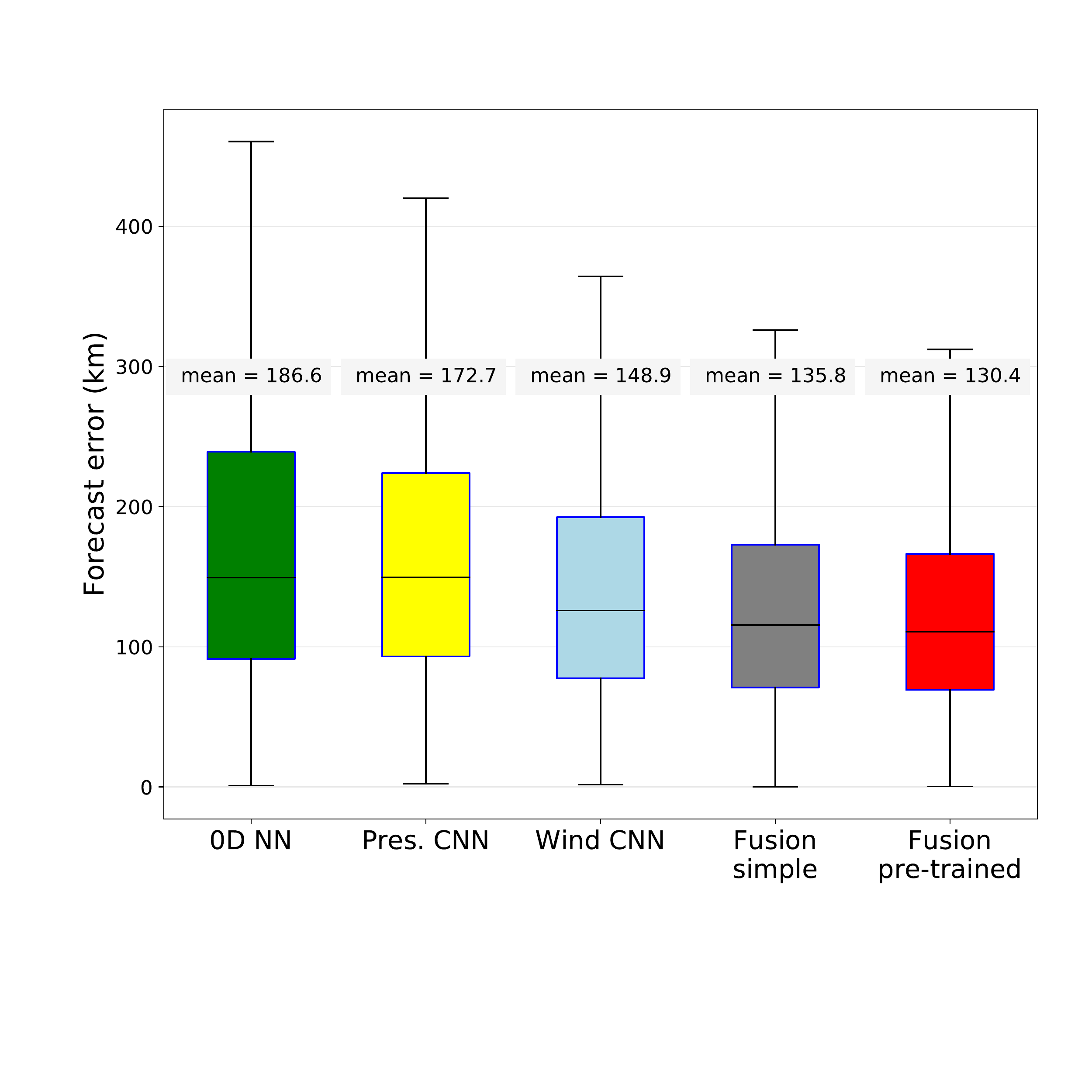}
	\caption{Comparison between the three simple networks (the 0D Neural Network, the Pressure CNN and the Wind CNN), the fused network without separate pre-training (gray), and the fused network with pre-training (red, proposed method). 24h-forecast results on the test set (storms coming from all oceanic basins), in distance between predicted and real locations.}
	\label{fig:boxplots}
\end{figure}

We have compared the fused network, fusing all three branches, with the three single branches of the networks. Figure \ref{fig:boxplots} shows the 24h-forecast results on the test set, which was 14,256 time steps in total, in absolute distance error. We can see the improvement of fusing networks (mean error : 130 km) with respect to the Wind CNN (mean: 148.9 km), the Pressure CNN (mean: 172.7 km) and the Past tracks + meta NN (mean: 186.6 km) alone. We can also see the importance of separately pre-training the three networks before the fusion, as it improves the mean result by 5 km.

We have also calculated a persistence forecast baseline: a 24-hour prediction that is four times the storm's last displacement from $t-6h$ to $t$. The mean error of this baseline on the test set is 196 km, which is more than 60 km higher than our method.

\begin{figure}[t]
  \centering
		\includegraphics[width=0.99\hsize]{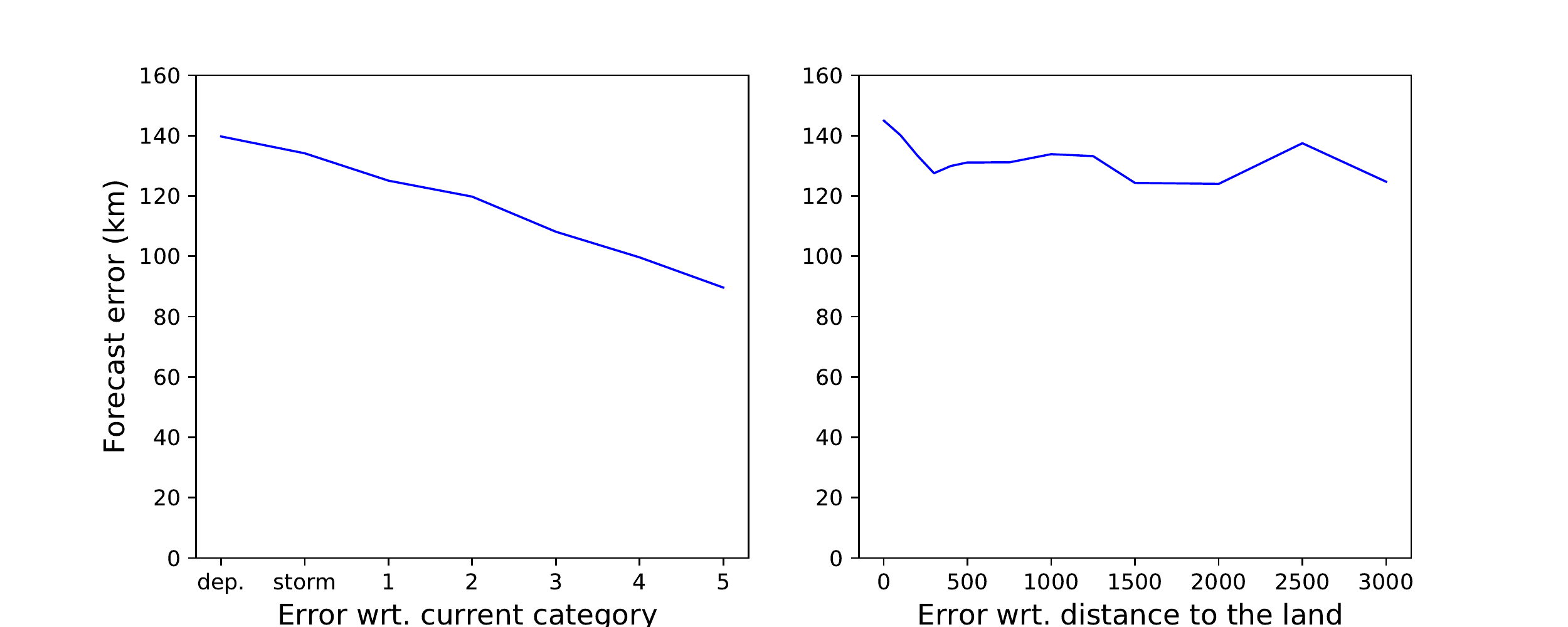}
	\caption{24h-forecast mean errors on the whole test set with respect to (a) the current Saffir-Simpson hurricane category (a higher category means a stronger hurricane, \textit{dep} means tropical depression, \textit{storm} means tropical storm); (b) its current distance to land. }
	\label{fig:res_cat_dist}
\end{figure}

Moreover, if we only test on tropical cyclone time steps excluding depressions, which are storms of lower intensity, our mean prediction error drops from 130 km to 109.3 km. Observe in Figure \ref{fig:res_cat_dist}(a) the global trend, showing that tracks from more intense storms are predicted with a lower mean error than less intense storms. The mean error from tropical cyclones of categories 4 and 5 is below 90km. Figure \ref{fig:res_cat_dist}(b) shows the forecast errors with respect to the current distance to land. We can see that a small distance to land, 200km or less, is one of the factors impacting the prediction quality. Lastly, we can see in Table \ref{tab:res_regions} for the results on the test set for the different regions or basins that the best results are in the North Atlantic with a mean error of 130.2km, or 26.5\% of the 24h displacement mean distance. The larger error is found in the South Pacific basin, but it is also the basin where we have the smaller number of samples. It is also worth notice that the IBTrACS database is made of different source agencies depending on the location of the storm, and their reliability are not equal: this might also explain why some basins show higher errors.
\\

\begin{table}[t]
\caption{24h-forecast results for the different regions (basins), on the test set. Mean error in km and relative mean error wrt. the mean 24h displacement distance.}\label{tab:res_regions}
\begin{center}
\begin{tabular}{ccccrrcrc}
\hline\hline
$Basin$ & mean error (km) & rel. to mean disp. (\%) & Num. time points\\
\hline
 North Atlantic  & 130.2 & 26.5\% & 2413 \\
 West Pacific & 136.1 & 27.9\% & 4080 \\
 East Pacific & 106.9 & 29.3\% & 2142 \\
 South Pacific & 161.7 & 41.6\% & 693 \\
 North Indian & 138.9 & 51.3\% & 2286\\
 South Indian  & 136.1 & 41.2\% & 3050\\
\hline
\end{tabular}
\end{center}
\end{table}

\subsection{Comparison with Statistical/Consensus Forecasts Methods}

We also compared our fusion model CNN with two existing forecasting models: CLP5\footnote{best track decay, combination of CLIPER5 (Climatology and Persistence model 5 day) and Decay-SHIFOR (Decay Statistical Hurricane Intensity Forecast).}, a statistical model which is often used to benchmark other storm track forecasting methods, and OFCL, the National Hurricane Center official forecast (consensus of dynamical models)\footnote{National Hurricane Center Forecast Verification, \url{https://www.nhc.noaa.gov/verification/verify6.shtml}, Accessed: 2018-07-31.}. We extracted the CLP5 prediction results of years 1989-2016 in the Atlantic and Eastern Pacific basins. We compare in Table \ref{tab:compare_with_art} our fused network with the statistical CLP5 on the test tropical cyclone time instants at which both methods provided a forecast. This means we compared only when there is a one-to-one correspondence, which is 4349 time steps from 258 storms. On both basins, our fused network performs better than the CLP5 model on average and in standard deviation. Moreover, the frequency of forecast errors larger to 200km, or \textit{busts}, is also lower for our method, especially in the Atlantic (10\% compared to 18\%). Such comparison is not possible with the OFCL as this model is modified every year and they only provide forecasts of the version N of the model for the year N. We don’t know the performance of the recent OFCL models on previous years and it would be unfair for them to compare with old results that were potentially obtained with earlier, less efficient models.

\begin{table}[]
	\caption{Mean and standard deviation 24h-forecast errors for the Atlantic and Pacific basins on the subset of the test set where both predictions were available (total = 4349 time steps). \textit{Busts} correspond to the ratio of track errors exceeding 200km (and 250km).}
	\label{tab:compare_with_art}
	\centering
	\small
	\renewcommand{\arraystretch}{2}
		\begin{tabular}{|c|c|c|c|c|c|c|c|c|}
			\hline
			\multirow{2}{*}{Model} & \multicolumn{4}{c|}{Atlantic errors (km)} & \multicolumn{4}{c|}{East Pacific errors (km)} \\ \cline{2-9} 
			& mean  & std & busts $>200km$ & busts $>250km$ & mean  & std & busts $>200km$ & busts $>250km$ \\ \hline
			CLP5 & 125 & 90 & 18.3\% & 9.8\% & 112 & 78 & 4.4\% & 2.3\% \\ \hline
			\textbf{Fusion} & 112 & 71 & 10.3\% & 4.5\% & 88 & 52 & 4.1\% & 1.9\% \\ \hline
		\end{tabular}%
\end{table}

\begin{figure}[t]
	\centering
		\includegraphics[width=1.1\hsize]{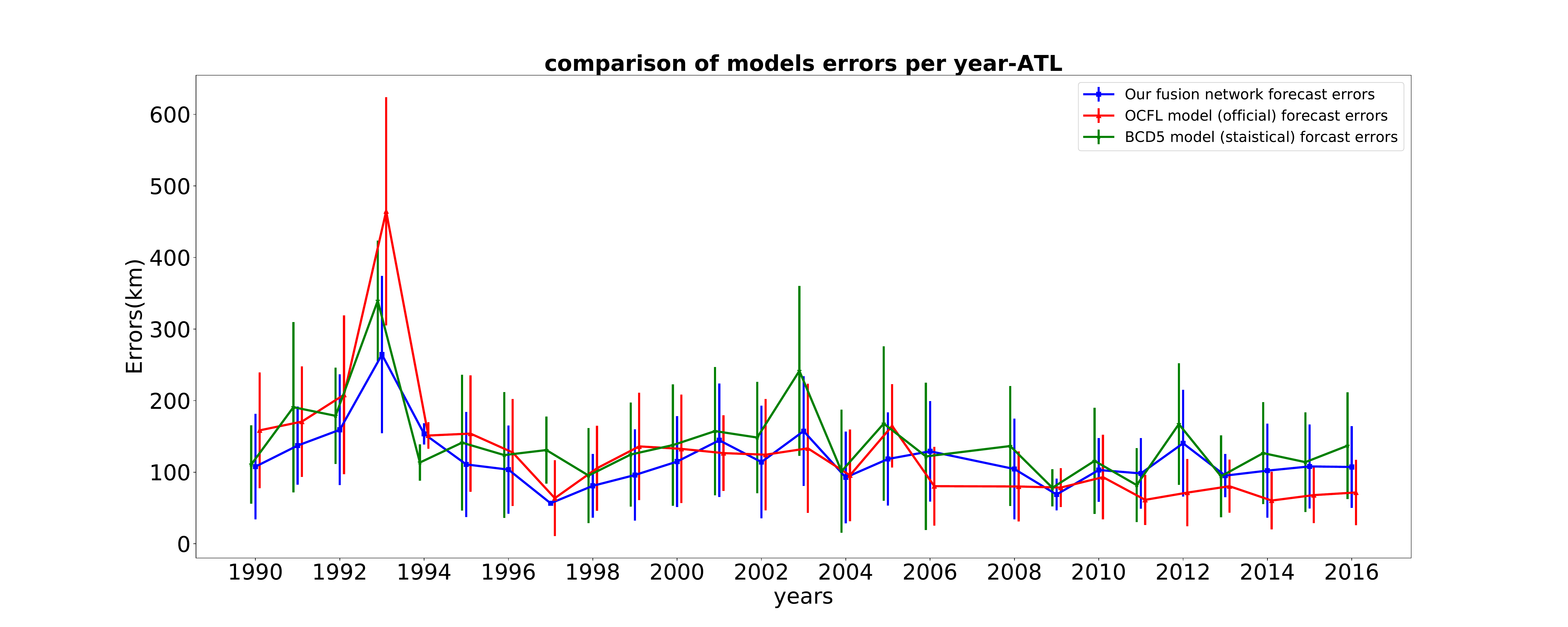}\\
		\includegraphics[width=1.1\hsize]{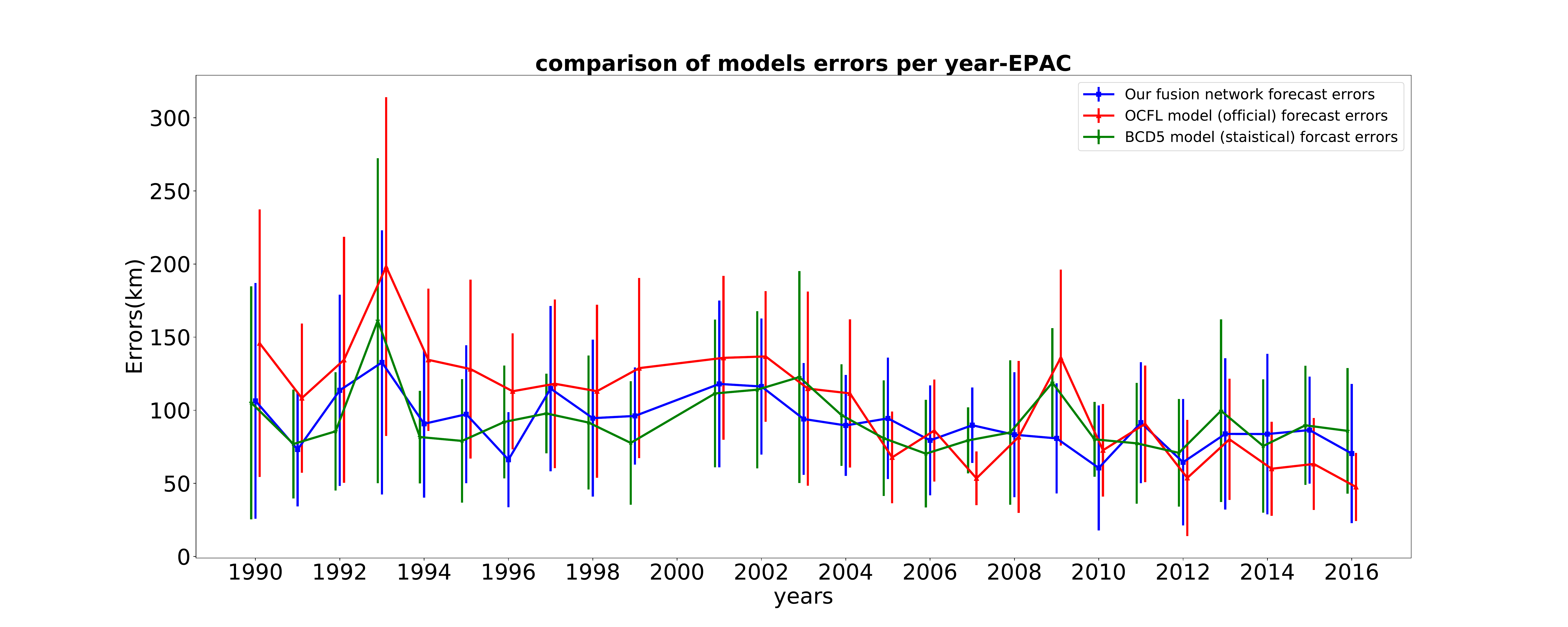}
	\caption{Yearly average of 24-hours storm track forecasting errors (km) and standard deviation on the test set (top figure for storms in Atlantic, bottom figure for storms in East Pacific) for our fused network forecasts (blue), the CLP5 model forecasts (green) and the official NHC forecasts (red), 1989-2016.}
	\label{fig:compare_with_art}
\end{figure}

\begin{figure}[t]
	\centering
		\includegraphics[width=1.1\hsize]{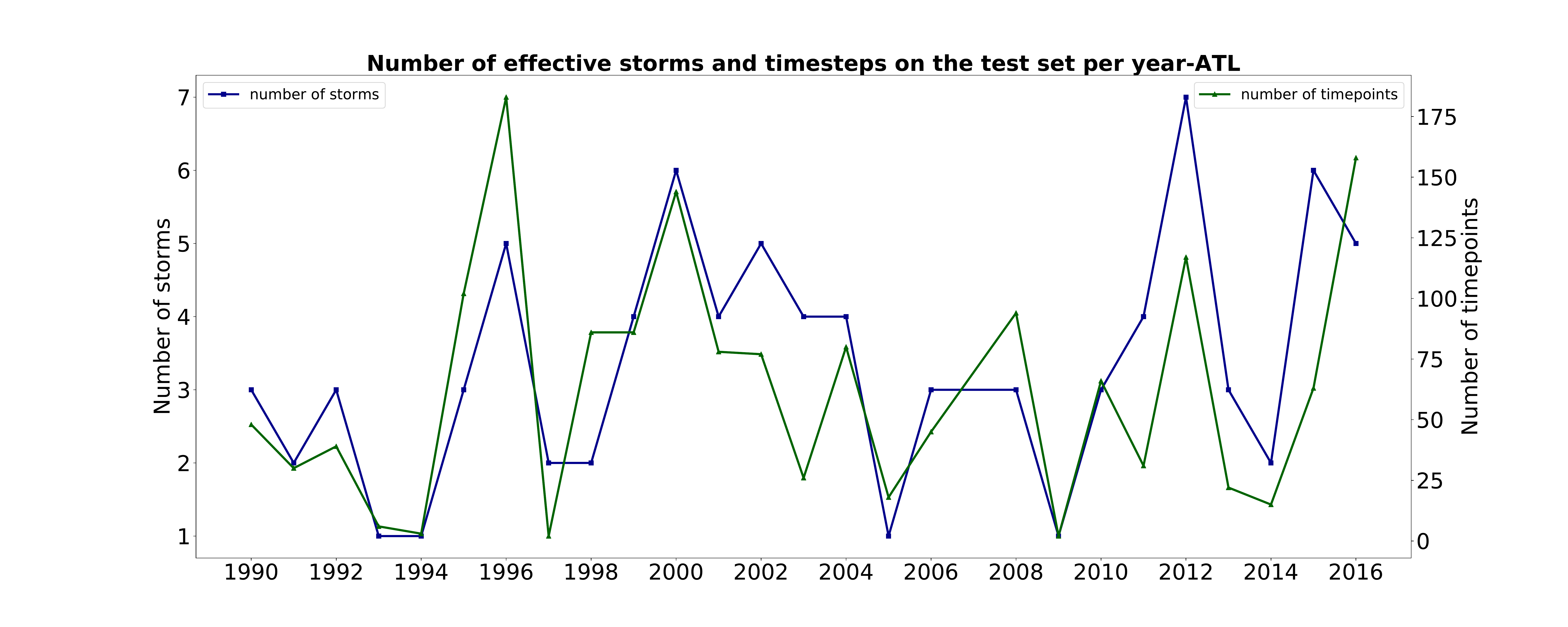}\\
		\includegraphics[width=1.1\hsize]{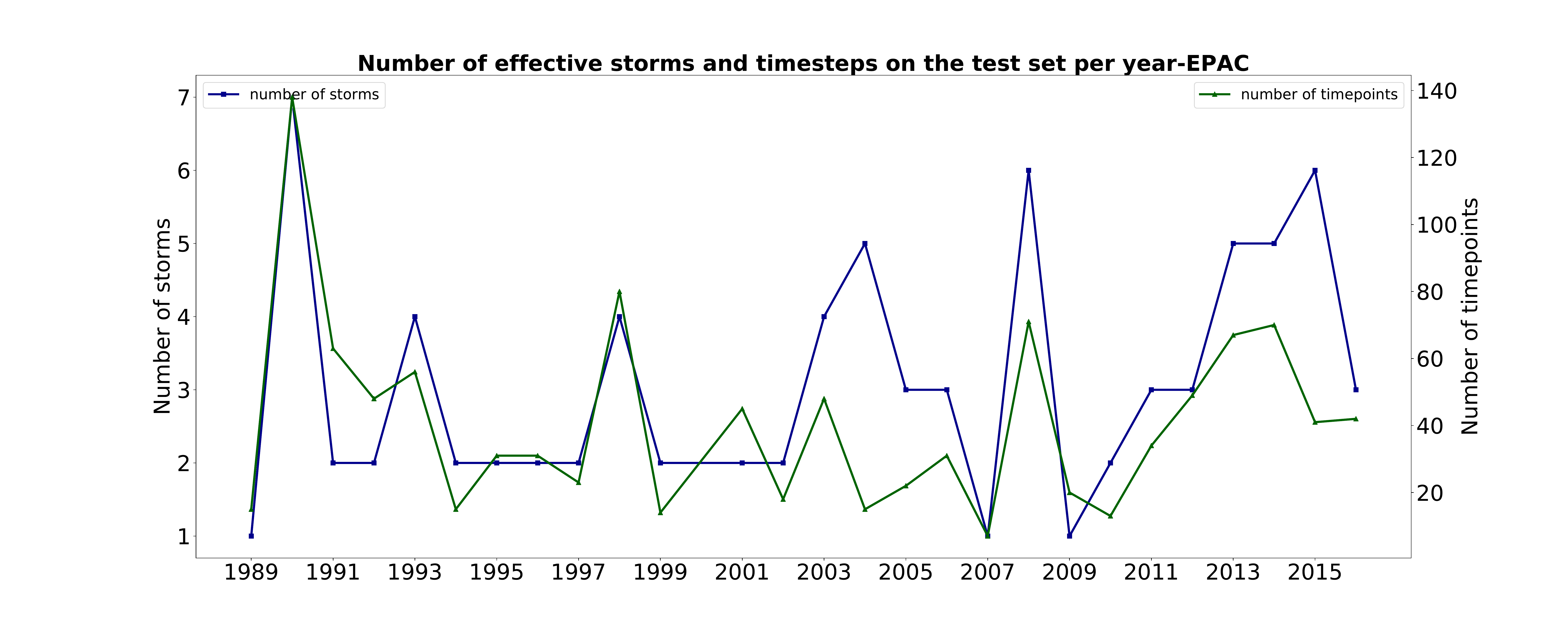}
	\caption{Number of storms and timesteps used to compare in the two basins (Atlantic and East Pacific) for every year, 1989-2016.}
	\label{fig:compare_number}
\end{figure}

That is why we compared the \emph{yearly} results of our fused network performance with the two models on the same subset of the test set. These results (mean and standard deviation) per year are shown in Figure \ref{fig:compare_with_art}. The number of storms and time steps each year for this comparison is presented in Figure \ref{fig:compare_number}. From the plot, we can see that our fused network behaves better than the statistical approach CLP5 in most of the years. Our deep learning model performs better than the OFCL forecast until year 2010 for the Pacific basin (2006 for the Atlantic). During the 2010s, the OFCL method improved and its mean errors per year were smaller than ours. We can also notice that none of the large error peaks (ATL: 1993, 2003, 2012; EPAC: 1993, 2009, 2013) involve our model, which seems to indicate that our method is robust.

\begin{figure}[h!]
  \centering
   \includegraphics[width=0.3\hsize]{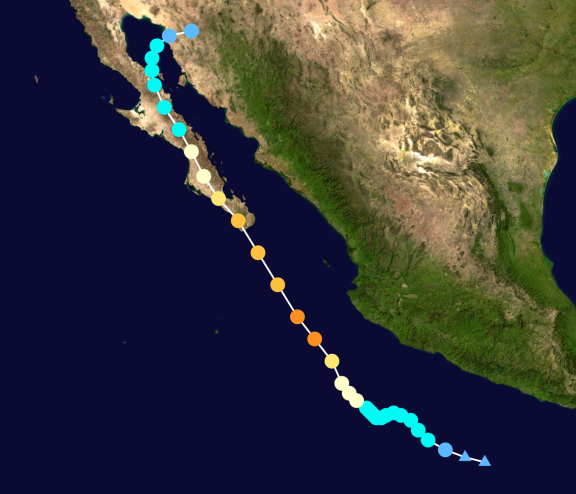}
	\includegraphics[width=0.9\hsize]{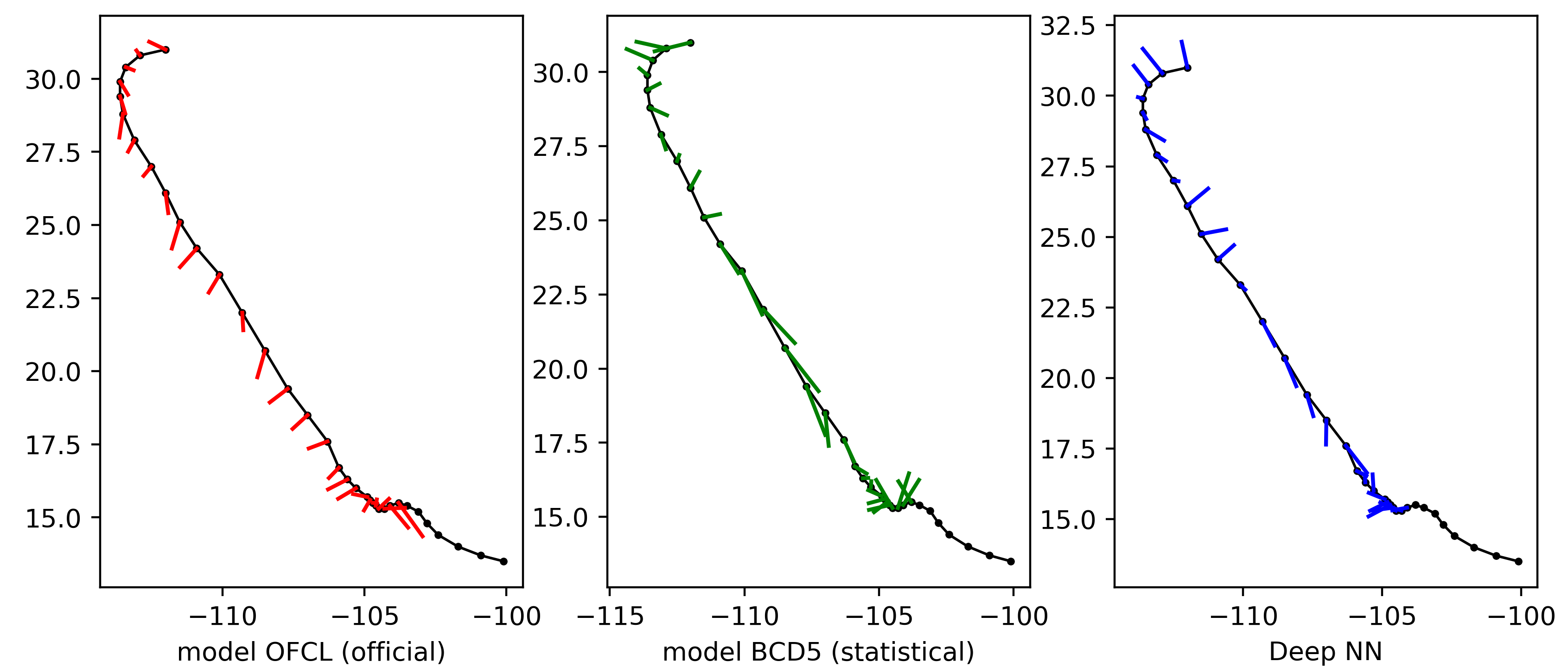}
  \caption{24-h forecast errors (4 time steps ahead) on Tropical Cyclone Odile in 2016. The bars connect each pair of predicted and ground truth location. The longer the length, the larger the error. At the beginning, the forecasts were not always available (a complete absence of an error bar should be interpreted as \textit{no forecast}).}
   \label{fig:odile}
\end{figure}

\begin{figure}[h!]
  \centering
   \includegraphics[width=0.3\hsize]{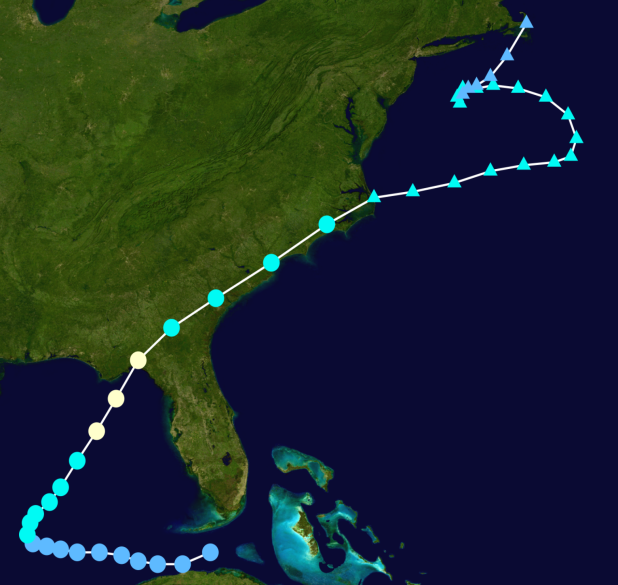}
\includegraphics[width=0.9\hsize]{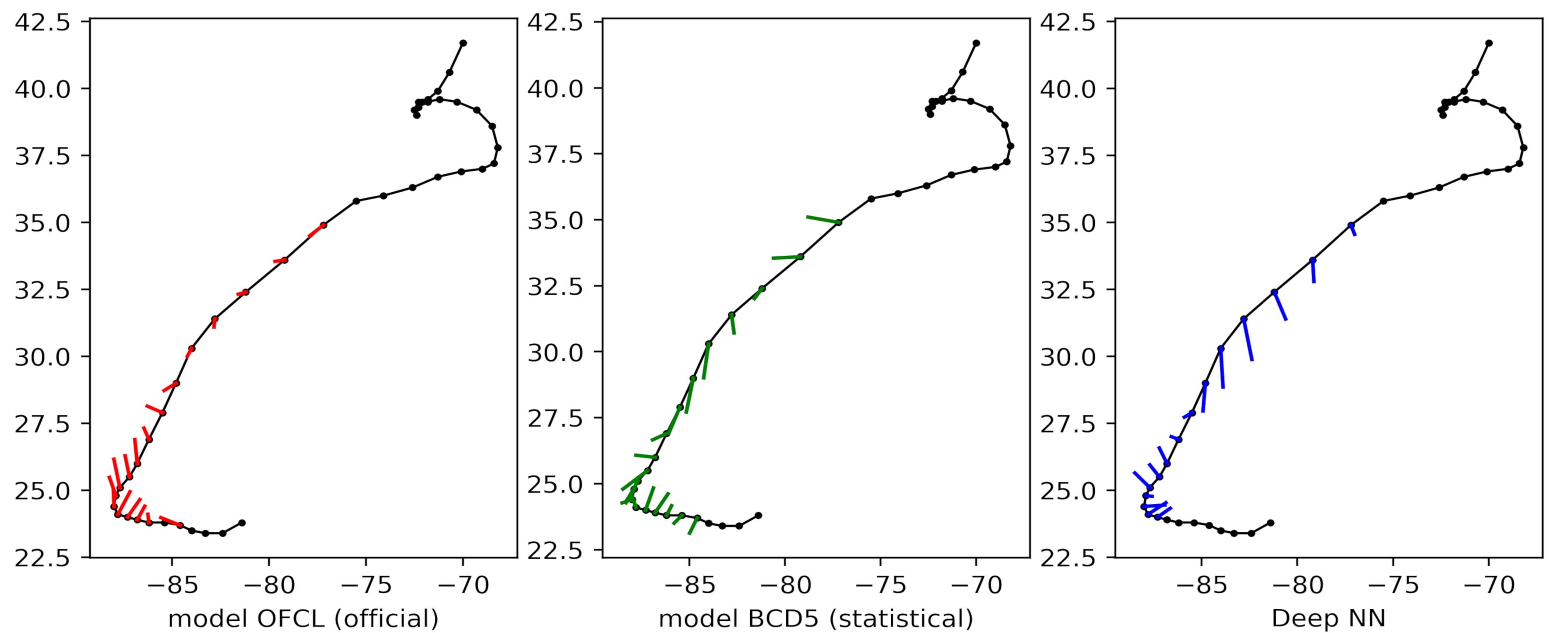}
  \caption{24-h forecast errors (4 time steps ahead) on Tropical Cyclone Hermine in 2016. The bars connect each pair of predicted and ground truth location. The longer the length, the larger the error. At the beginning and at the end of the track, the forecasts were not always available (a complete absence of an error bar should be interpreted as \textit{no forecast}).}
   \label{fig:hermine}
\end{figure}

\begin{figure}[h!]
  \centering
 \includegraphics[width=0.4\hsize]{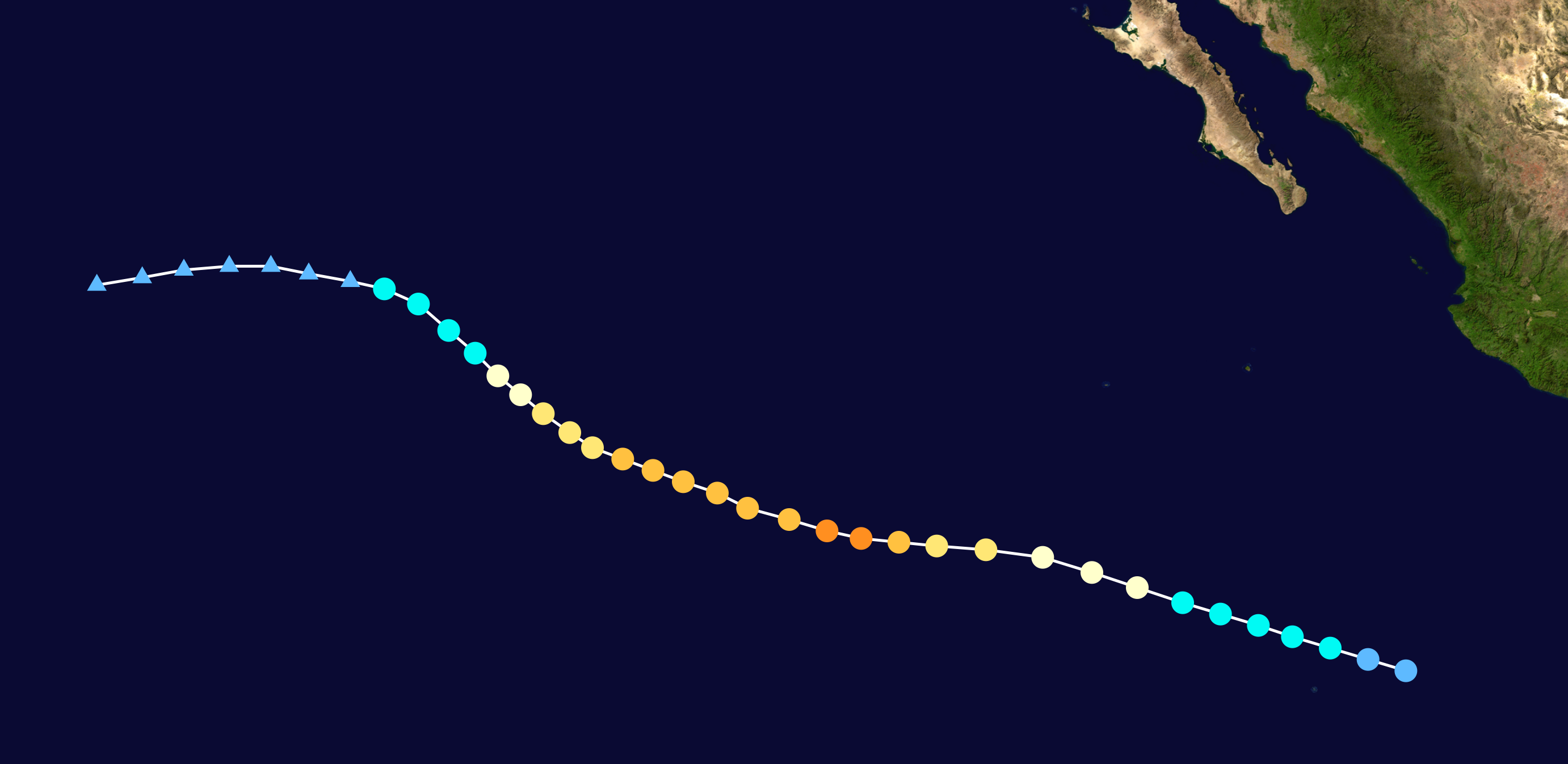}
  \includegraphics[width=0.9\hsize]{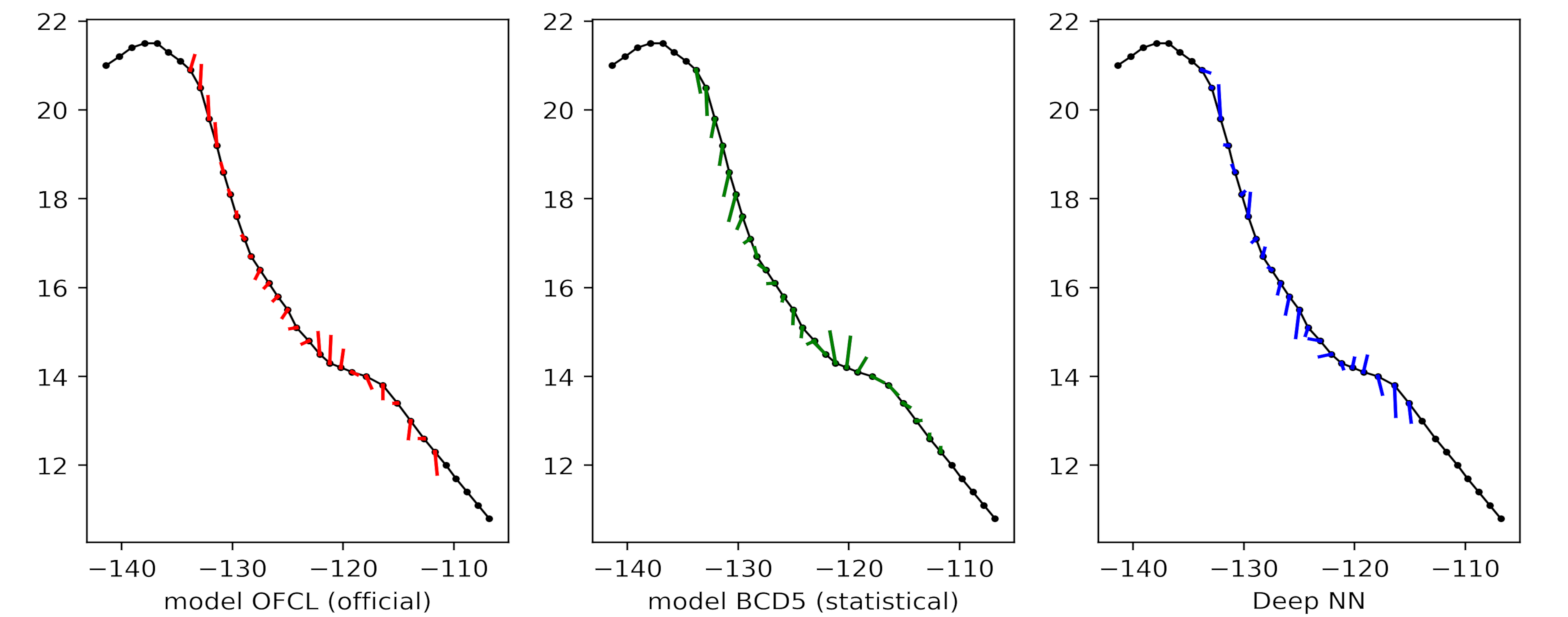}
  \caption{24-h forecast errors (4 time steps ahead) on Tropical Cyclone Blas in 2016. The bars connect each pair of predicted and ground truth location. The longer the length, the larger the error. At the beginning and at the end of the track, the forecasts were not always available (a complete absence of an error bar should be interpreted as \textit{no forecast}).}
   \label{fig:blas}
\end{figure}

Finally, we qualitatively compared the predictions with both OFCL and CLP5 models for recent storms of the test set, such as Tropical Cyclone Odile in 2014 (Figure \ref{fig:odile}), Tropical Cyclone Hermine in 2016 (Figure \ref{fig:hermine}), and Tropical Cyclone Blas in 2016 (Figure \ref{fig:blas}). The small bars connect each pair of predicted and ground truth location after 24 hours. The longer the length, the larger the error. Even though the official OFCL model has globally smaller forecast errors, on some time points our model outperforms the OFCL. It seems that our method still perform poorly on the land (see Figure \ref{fig:hermine}). A future improvement could be to add the sea/land map as additional feature. Moreover, the three forecasts often have different directions. Quantitatively, we calculated the Pearson correlation coefficient between the errors of the proposed method and the baselines: $c_{fusion, OFCL}=0.33$, $c_{fusion, CLP5}=0.55$, 
 also suggesting that the methods behave very differently. A neural network model can thus help the current forecast modellers by providing a complementary prediction that could be integrated in a consensus method, as their mistakes are different.
\\

\section{Discussion}

Our method only needs to be trained once, although this training can be improved with more data. After that, only a few seconds are needed to give a forecast for a new storm because prediction or inference using such models is much faster than training them. This is a significant time improvement over dynamical models, whose bottleneck is the computing speed. However, one has to keep in mind that our method needs current and past reanalysis fields. While they are usually quickly calculated, within few hours, it does increase the total forecast time accordingly.

We have shown a proof-of-concept for 24-hour forecasting, and \citet{giffardroisin:hal-01851001} shows that the 6-hour results are also very satisfactory. Yet, more long-term forecasts could be made using the same structure. We conjecture that for very long forecasts, larger than 25x25 degree images might be needed. Moreover, we worked here on trajectory prediction, yet this model can be easily modified by changing the last layer and be trained for another task, such as intensity prediction (see \citet{giffard20182018}). The data was limited to the period 1979-2017, but more data will be acquired each year and the method can improve significantly by adding the new storms in the training set, as for now we still have fewer samples than parameters. With new TC data available each year and with larger image sizes, it will be necessary to scale our method to a larger number of GPUs and make full profit of data parallelism techniques \citep{krizhevsky2014one}.

Other useful features could be found by using different reanalysis fields. Although our choice of wind and geopotential height fields was driven by an automated feature selection method, we did not test all the possible configurations at every pressure level. Potentially, a more refined selection could increase the overall performance. As an example, for the intensity prediction, we think that surface fields such as sea surface temperature should be reconsidered. Moreover, as it was shown that the storms close to the land or on the land have higher forecast errors, we think that some improvements can be made by adding the land-sea mask image to the 2D features. We could also represent the wind field by streamfunction and velocity potential as opposed to u- and v-wind components, which might help to have less correlated features. Moreover, while the machine learning algorithm could learn the differences of flow direction between North and South, a future improvement could be to flip the fields North-South and to change the sign of the vwind component. The recent release of the new version of ERA reanalysis, ERA 5, might also increase the accuracy. As \citet{hodges2017well} show, the mean offset in tropical cyclone center position in the ERA-Interm reanalysis product can be up to 1 degree for the period from 1979 to 2012, so moving to ERA 5 and using the GFDL Vortex Tracker \citep{marchok2002ncep} would increase our performance. A comparison to other baseline forecasts, such as TVCN (Track Variable ConseNsus), would also be interesting. Finally, our method could be easily transferred to operational Numerical Weather Prediction data by filtering it to the same spatial resolution.
\\

\section{Conclusion}

We designed a neural network for the 24h-cyclone storm track forecasting using a moving frame of reference that makes use of a common dataset and a unique trained NN for every tropical cyclone of both hemispheres. When a new tropical cyclone occurs, our network can give a forecast in only few seconds. We demonstrated the benefit of coupling past displacements and aligned reanalysis images. Moreover, we also compared with traditional forecasting methods and showed the improvement with respect to the statistical CLP5 model. This is only a proof-of-concept of deep learning for tropical cyclone forecasting, yet we think that such a different approach as machine learning and NN can be very beneficial if integrated in a consensus method.
\\

\section*{Conflict of Interest Statement}

The authors declare that the research was conducted in the absence of any commercial or financial relationships that could be construed as a potential conflict of interest.

\section*{Author Contributions}

S.G., M.Y., C.M. and G.C. have designed the architecture; S.G. and M.Y. have developed the codes; G.C. and B.K. have provided the infrastructure to develop and train the models; C.K. has proof-read the manuscript.

\section*{Funding}
The research leading to these results has received funding from the Jean D'Alembert fellowship from the Fondation Campus Paris-Saclay and DirtyData (ANR-17-CE23-0018) grant.

\section*{Acknowledgments}
We thank the National Oceanic and Atmospheric Administration (NOAA) and the European Centre for Medium-Range Weather Forecasts (ECMWF) for their online databases IBTrACS and ERA-Interim. 
This manuscript has been released as a Pre-Print at arXiv \citep{giffardroisin2019tropical}.


\bibliographystyle{frontiersinSCNS_ENG_HUMS} 
\bibliography{references}

\end{document}